\documentclass[final]{siamltex}
\usepackage{amsmath}
\usepackage{amssymb}
\usepackage[ruled,section]{algorithm}
\usepackage{algorithmic}

\usepackage[square,comma,numbers,sort&compress]{natbib} 

\usepackage{ifpdf}
\ifpdf 
\usepackage[pdftex]{graphicx}      
\DeclareGraphicsExtensions{.pdf}   
\usepackage[pdftex,                
pdfstartview=FitH,%
bookmarks=true,
bookmarksnumbered=true,
bookmarksopen=true,%
hypertexnames=false,
breaklinks=true,
  colorlinks=true,%
  linkcolor=blue,anchorcolor=blue,%
  citecolor=blue,filecolor=blue,%
  menucolor=blue]%
{hyperref} 
\hypersetup{
pdfauthor   = {A. V. KNYAZEV, M. E. ARGENTATI, I. LASHUK, AND E. E. OVTCHINNIKOV},
pdftitle    = {Block Locally Optimal Preconditioned Eigenvalue Xolvers (BLOPEX) 
in hypre and PETSc},
pdfsubject  = {Paper},
pdfkeywords = {Conjugate gradient, 
iterative method, 
preconditioning, 
preconditioner,
multigrid, 
domain decomposition, 
parallel computing, 
eigenvalue, 
eigenvector, 
Laplacian, 
LOBPCG, 
BLOPEX, 
hypre, 
PETSc, 
BlueGene, 
Beowulf. },
}
\else 
\usepackage[dvips]{graphicx}       
\DeclareGraphicsExtensions{.eps}   
\usepackage[ps2pdf]{thumbpdf}      
\usepackage[ps2pdf,                
hypertexnames=false,
breaklinks=true,
  colorlinks=true,%
  linkcolor=blue,anchorcolor=blue,%
  citecolor=blue,filecolor=blue,%
  menucolor=blue]%
{hyperref} 
\hypersetup{
pdfauthor   = {A. V. KNYAZEV, M. E. ARGENTATI, I. LASHUK, AND E. E. OVTCHINNIKOV},
pdftitle    = {Block Locally Optimal Preconditioned Eigenvalue Xolvers (BLOPEX) 
in hypre and PETSc},
pdfsubject  = {Paper},
pdfkeywords = {Conjugate gradient, 
iterative method, 
preconditioning, 
preconditioner,
multigrid, 
domain decomposition, 
parallel computing, 
eigenvalue, 
eigenvector, 
Laplacian, 
LOBPCG, 
BLOPEX, 
hypre, 
PETSc, 
BlueGene, 
Beowulf.},
pdfcreator  = {LaTeX with hyperref package},
pdfproducer = {dvips + ps2pdf}
}
\fi 

\def\betab{\vspace{-10pt}\begin{tabbing}
x\=xxx\=xx\=xxx\=xxx\=xxx\=xxx\=xxx\=xxx\= \kill}
\def\entab{\end{tabbing}\vspace{-0.12in}}

\title{Block Locally Optimal Preconditioned Eigenvalue Xolvers (BLOPEX) 
in {\it hypre} and PETSc
\thanks{
Received by the editors May 31, 2006; 
accepted for publication (in revised form) ????????????;
published electronically \today.
Preliminary results of this work have been presented at:   
11th and 12th Copper Mountain Conferences on Multigrid Methods, 
Preconditioning 2003, 
SIAM Parallel Processing for Scientific Computing 2004, 
and 16th International Conference on Domain Decomposition Methods  
and appeared in \citet{ka05tr,iaok06}. 
}}
\author{
A. V. Knyazev\footnotemark[2]\ \footnotemark[3]
\and M. E. Argentati\footnotemark[2]\ \footnotemark[4] 
\and I. Lashuk\footnotemark[2]\ \footnotemark[5]
\and  E. E. Ovtchinnikov\footnotemark[6]
}
\begin{document}
\setcounter{page}{1}
\maketitle
\renewcommand{\thefootnote}{\fnsymbol{footnote}}
\footnotetext[2]{Department of Mathematical Sciences, 
University of Colorado at Denver and Health Sciences Center,
P.O. Box 173364, Campus Box 170, Denver, CO 80217-3364, USA.
Partially supported by NSF DMS 0208773 and 0612751,
and by the 
Lawrence Livermore National Laboratory.}
\footnotetext[3]{{\tt Andrew.Knyazev[at]cudenver.edu, http://math.cudenver.edu/\~{}aknyazev/}}
\footnotetext[4]{{\tt Merico.Argentati[at]cudenver.edu, http://math.cudenver.edu/\~{}rargenta/}}
\footnotetext[5]{{\tt Ilya.Lashuk[at]cudenver.edu}}
\footnotetext[6]{Harrow School of Computer Science, 
University of Westminster, London HA1 3TP, UK. 
{\tt Evgueni.Ovtchinnikov[at]na-net.ornl.gov}}

\begin{abstract}
We describe our software package Block Locally Optimal Preconditioned Eigenvalue Xolvers (BLOPEX) 
publicly released recently. BLOPEX is available 
as a stand-alone serial library, as an external package 
to PETSc (``Portable, Extensible Toolkit for Scientific Computation'',
a general purpose suite of tools for the scalable solution 
of partial differential equations and related problems developed by Argonne National Laboratory), 
and is also built into {\it hypre} (``High Performance Preconditioners'',
scalable linear solvers package developed by Lawrence Livermore National Laboratory).
The present BLOPEX release includes only one 
solver---the Locally Optimal Block Preconditioned Conjugate Gradient (LOBPCG) 
method for symmetric eigenvalue problems. {\it hypre} provides users 
with advanced high-quality parallel preconditioners for linear systems, 
in particular, with domain decomposition and multigrid preconditioners.  
With BLOPEX, the same preconditioners can now be efficiently used for 
symmetric eigenvalue problems. PETSc facilitates the integration of independently developed 
application modules with strict attention to component interoperability, and makes 
BLOPEX extremely easy to compile and use with preconditioners that are available via PETSc.
We present the LOBPCG algorithm in BLOPEX for {\it hypre} and PETSc. 
We demonstrate numerically the scalability of BLOPEX by testing it on a number of
distributed and shared memory parallel systems, including
a Beowulf system, SUN Fire 880, an AMD dual-core Opteron workstation, 
and IBM BlueGene/L supercomputer, using PETSc domain decomposition and {\it hypre}
multigrid preconditioning. We test BLOPEX on a model problem, the standard 7-point 
finite-difference approximation of the 3-D Laplacian, with the problem size in the range $10^5-10^8$. 
\end{abstract}
\begin{AM}
65F15 
65N25 
65N55 
65Y05 
\end{AM}

(Place for Digital Object Identifier, to get an idea of the final spacing.) 
\pagestyle{myheadings}
\thispagestyle{plain}
\markboth{A. KNYAZEV, M. ARGENTATI, I. LASHUK, AND E. OVTCHINNIKOV}
{BLOPEX IN HYPRE AND PETSC}
\section{Introduction}
We describe a new software package 
Block Locally Optimal Preconditioned Eigenvalue Xolvers (BLOPEX) 
revision 1.0 
publicly released recently. 
BLOPEX is available 
as a stand-alone serial library, which can be downloaded from 
{\tt http://math.cudenver.edu/$\,\tilde{}\,$aknyazev/software/BLOPEX/},
as an external package 
to the Portable, Extensible Toolkit for
Scientific Computation (PETSc) \cite{petsc-user-ref}, 
and is built into the High Performance Preconditioners ({\it hypre})
library \citet{Falgout:2005:PSH,Falgout:2006}. 
Jose Roman has also written a  
Scalable Library for Eigenvalue Problem Computations (SLEPc)
\cite{Hernandez:2005:SSF})
interface to our {\it hypre} BLOPEX. 

The native {\it hypre} BLOPEX version efficiently and directly 
takes advantage of powerful {\it hypre} multigrid preconditioners, both  
algebraic (AMG, called BoomerAMG in {\it hypre}) and geometric or structured (SMG).
BLOPEX in PETSc gives the PETSc community easy access to the
customizable code of a preconditioned eigensolver.

At present, BLOPEX includes only one 
solver---the Locally Optimal Block Preconditioned Conjugate Gradient (LOBPCG)
\citet{k99,k00} method for eigenproblems  
$A x = \lambda B x$
with large, possibly sparse, symmetric matrix $A$ 
and symmetric positive definite matrix $B$.
Preconditioned eigenvalue solvers in general, see 
\citet{k98,k99}, and in particular
the LOBPCG method have recently attracted 
attention as potential competitors to (block-) Lanczos methods.  
LOBPCG has been 
implemented using different computer languages in a number of other software packages: 
C++ in Anasazi Trilinos \citet{ahlt05,tri05} and 
in NGSolve 
\citet[Sections 7.2-7.4]{sz_phd_06}; 
C in PRIMME \citet{primme06}; 
FORTRAN 77 in PLTMG \citet{PLTMG90};  
FORTRAN 90 in ABINIT \citet{blkz07} 
and in PESCAN \citet{tlcwd05}; 
Python in SciPy by Robert Cimrman and in PYFEMax by Peter Arbenz and Roman Geus. 
LOBPCG has been independently tested in
\citet{1105808,yikm06} for Fermion-Habbard Model on Earth Simulator CDIR/MPI; 
in \citet{ahlt05} and \citet{MR2199542} using AMG preconditioning
for vibration problems;      
and in \citet{tlcwd05,ymw06} for electronic structure calculations. 
Our {\it hypre} BLOPEX version has been used in \citet{MR2164087} 
for the {M}axwell problem.

Section \ref{slobpcg} contains a complete
and detailed description of the LOBPCG algorithm as implemented in BLOPEX.
We discuss our specific abstract implementation approach in section \ref{sabs}.
Parallel performance on distributed and shared memory systems 
using domain decomposition and multigrid preconditioning is discussed in 
section \ref{snum}. 
We collect technical information, e.g.,\ the list of acronyms,  
in the appendix. 
\section{LOBPCG and its Implementations} 
\label{slobpcg}
In subsection \ref{slobpcg1}, we briefly describe the ideas behind the LOBPCG method
mostly following \citet{k99,k00}. 
In subsection \ref{slobpcg2}, we present a 
complete description of the LOBPCG algorithm as 
implemented in BLOPEX. 
Deflation by hard and soft locking is suggested in subsection \ref{slobpcg4}.
\subsection{LOBPCG Ideas and Description} 
\label{slobpcg1}
\subsubsection{The problem and the assumptions}
We consider the problem of computing the $m$ smallest 
eigenvalues and the corresponding eigenvectors of 
the generalized eigenvalue problem $A x = \lambda B x$, with
symmetric (Hermitian in the complex case) matrices $A$ and $B$,
the latter assumed to be positive definite. 
For a given approximate eigenvector $x$, 
we use the traditional Rayleigh quotient 
$\lambda (x) = (x,Ax)/(x,Bx)$ 
in the standard scalar product
as an approximation to the corresponding eigenvalue.
We emphasize that only matrix $B$ is assumed to be positive definite. 
There is no such requirement on $A$; moreover, 
in all formulas in this section we can replace $A$ with $A+\alpha B$ and 
the formulas are invariant with respect to a real shift $\alpha.$ 

To accelerate the convergence, we introduce 
a preconditioner $T$, which is typically a 
function that for a given vector $x$ produces 
$Tx$. 
The existing theory, e.g.,\ \citet{k00,kn03},
of the LOBPCG method is based on several assumptions: $T$ 
needs to be linear, symmetric, and positive definite. 
Under these assumptions a specific convergence rate bound  
is proved in \cite{k00,kn03}. 
Numerical tests suggest that 
{\em all these requirements on $T$ are necessary}  
in order to guarantee this convergence rate. 
The use of preconditioning that does not satisfy 
the requirements above not only can slow down 
the convergence, but may also lead to 
a breakdown of the method and severe instabilities 
in the code resulting in incorrect results---that should not be mistakenly interpreted 
as bugs in the code. The method is robust however if $T$ 
changes from iteration to iteration as soon as in every step it satisfies the above assumptions.  
\subsubsection{Single-Vector LOBPCG}
For computing only the smallest eigenpair, i.e. if $m=1$,  
the LOBPCG method takes the form of 
a 3-term recurrence:
\begin{equation}
\begin{tabular}{l}
$x^{(i+1)} = w^{(i)} + \tau^{(i)} x^{(i)} { + \gamma^{(i)} x^{(i-1)}}, $ \\
$w^{(i)} =  T(A x^{(i)} - \lambda^{(i)}  Bx^{(i)}), \
\lambda^{(i)} = \lambda (x^{(i)})
= (x^{(i)},A x^{(i)}) / (Bx^{(i)}, x^{(i)}) $ \\
\end{tabular} 
\label{old}
\end{equation}
with properly chosen scalar iteration parameters
$\tau^{(i)},$ and $\gamma^{(i)}$. 
The easiest and perhaps the most efficient choice
of parameters is based on an idea of 
{\em local optimality} \citet{k90,k99,k00}, 
namely, select $\tau^{(i)}$ and $\gamma^{(i)}$
that minimize the Rayleigh
quotient $\lambda(x^{(i+1)})$ by using the 
Rayleigh-Ritz procedure in 
the three-dimensional trial subspace
spanning $x^{(i)}$, $w^{(i)},$ and $x^{(i-1)}$.
We do not describe the Rayleigh-Ritz procedure
using an arbitrary basis of the trial subspace 
in this paper and assume that the reader is familiar with it.

If $\gamma^{(i)}=0$, method (\ref{old}) turns into 
the classical preconditioned locally optimal steepest 
descent method, so one can think of method (\ref{old}) 
as an attempt to accelerate the steepest 
descent method by adding an extra term to 
the iterative recurrence; e.g.,\ \citet{k90,k99}. 
The general idea of accelerating an iterative 
sequence by involving the previous approximation 
is very natural and is well known. Let us note here that 
if instead of the previous approximation $x^{(i-1)}$
we use the previous preconditioned residual $w^{(i-1)}$ 
in formula (\ref{old}), we get a different method that 
in practical tests converges not as fast as 
method (\ref{old}). 

When implementing the Rayleigh-Ritz procedure 
in the trial subspace 
spanning $x^{(i)}$, $w^{(i)},$ and $x^{(i-1)}$, 
we need to be careful in choosing an appropriate basis 
that leads to reasonably conditioned Gram matrices. 
The current eigenvector approximation $ x^{(i)}$ and 
the previous eigenvector approximation $ x^{(i-1)}$ get
closer to each other in the process of iterations, so 
special measures need to be taken in method (\ref{old}) 
to overcome the potential numerical instability 
due to the round-off errors. 
Using both vectors $x^{(i)}$ and $ x^{(i-1)}$  
as basis vectors of the trial subspace 
leads to ill-conditioned Gram matrices,  
and the Rayleigh-Ritz method can produce spurious eigenpairs.

A simple fix is suggested in \citet{k00}: 
the three-term recurrence that contains
the current eigenvector approximation, the 
preconditioned residual, and the implicitly computed  
difference between the current and the previous eigenvector approximations:
\begin{equation}
\begin{tabular}{rclcl}
$x^{(i+1)} $&$=$& $w^{(i)} + \tau^{(i)} x^{(i)}$ &$+$& $\gamma^{(i)} p^{(i)}, \ 
w^{(i)} = T (A x^{(i)} - \lambda^{(i)} B x^{(i)}),$ \\
$p^{(i+1)} $&$=$&$ w^{(i)}$ &$+$ & $\gamma^{(i)} p^{(i)}, \ p^{(0)} = 0, 
\ \lambda^{(i)} = \lambda (x^{(i)})$.  \end{tabular} 
\label{prec_cgmBA}
\end{equation}
$
x^{(i+1)} \in {\rm span} \{ w^{(i)}, \ x^{(i)}, \ p^{(i)} \}
=             {\rm span} \{ w^{(i)}, \ x^{(i)}, \ x^{(i-1)} \}
$
as $p^{(i+1)} = x^{(i+1)} - \tau^{(i)} x^{(i)},$ 
therefore, the new formula (\ref{prec_cgmBA})
is mathematically equivalent to (\ref{old}), in exact arithmetic. 
We choose in (\ref{prec_cgmBA}) the scalar iteration parameters
$\tau^{(i)}$ and $\gamma^{(i)}$  as above, i.e.
minimizing the Rayleigh quotient $\lambda(x^{(i+1)})$. 

We note that formula (\ref{prec_cgmBA}) is not quite 
the ultimate fix, e.g.,\ 
it is possible in principle that 
at the initial stage of the iterations  
$\tau^{(i)}$ is small, so that $p^{(i+1)}$  
is too close to $x^{(i+1)}$, 
and the Rayleigh-Ritz procedure may fail.
The proper choice of the basis in the 
trial subspace ${\rm span} \{w^{(i)}, x^{(i)}, p^{(i)} \}$
for the Rayleigh-Ritz procedure in (\ref{prec_cgmBA}), 
having in mind that vectors $w^{(i)}$ and $p^{(i)}$ 
converge to zero, is not a trivial issue, see 
\citet{hl06} for possible causes of LOBPCG instability.

The locally optimal choice of the 
step sizes in the LOBPCG allows an easy generalization  
from the single-vector version  (\ref{old}) or (\ref{prec_cgmBA}) 
to the block version described next in subsection \ref{ss:bas}, 
where a block of $m$ vectors is iterated simultaneously. 
We return to the discussion 
of the choice of the basis in subsection \ref{slobpcg2}, where it
becomes even more vital with the increase of the 
block size $m$. 
\subsubsection{The LOBPCG Basics}\label{ss:bas}
A block version of LOBPCG for finding the $m\geq 1$ smallest eigenpairs  
is suggested in \citet{k98,k99}:
\begin{equation*}  
x^{(i+1)}_j \in {{\rm span}}
\left\{ x^{(i-1)}_1, \ x^{(i)}_1, \ T (A- \lambda^{(i)}_1 B) x^{(i)}_1, \ldots ,
\ x^{(i-1)}_m, \ x^{(i)}_m, \ T (A- \lambda^{(i)}_m B) x^{(i)}_m \right\},
\end{equation*}
where $x^{(i+1)}_j$ is computed as the $j$-th Ritz vector, 
$j=1, \ldots, m$ corresponding to the $j$-th smallest Ritz value in 
the Rayleigh-Ritz procedure 
on a $3m$-dimensional trial subspace. The block method has the same problem 
of having close vectors in the trial subspace
as the single-vector version (\ref{old}) discussed in the 
previous subsection,
for which \citet{k00} suggested the same fix using the  
directions $p$. 

As in other block methods, the block size should be chosen,
if possible, to provide a large gap between first $m$
eigenvalues and the rest of the spectrum as this typically
leads to a better convergence, see \citet{k00,kn03,ovtch06}.
Block methods generally handle 
clusters in the spectrum and multiple eigenvalues quite well;
and the LOBPCG is no exception. An attempt should be
made to include the whole cluster of eigenvalues into the 
block, while for multiple eigenvalues this is not essential at all. 
The numerical experience is that the smallest eigenvalues 
often converge faster, and that the 
few eigenvalues with the noticeably slow convergence 
usually are the largest in the block and typically 
correspond to a cluster of eigenvalues that is not 
completely included in the block. 
A possible cure is to use a flexible block size a bit 
larger than the number of eigenpairs actually wanted  
and to ignore poorly converging eigenpairs. 

The block size $m$ should not be so big that  
the costs of the  Rayleigh-Ritz procedure 
in the 
$3m$-dimensional trial subspace dominate the 
costs of iterations. 
If a large number of 
eigenpairs is needed, deflation is necessary, 
see subsection \ref{slobpcg4}.   
The optimal choice of the block size $m$ 
is problem-, software- and computer-dependent; 
e.g.,\ it is affected by such details as the amount of the CPU cache    
and the use of optimized BLAS library functions in the multivector implementation, 
see section \ref{sabs}.
\subsection
{The Detailed Description of the LOBPCG Algorithm} \label{slobpcg2}
The description of the LOBPCG
algorithm as implemented in our  
BLOPEX-1.0 code follows: \\
{\betab
{\underline {\bf Input:}} $m$ starting linearly independent vectors in
$X \in  
\mathbb{R}^{n \times m}$, 
$l$ linearly independent\\ 
\>\> constraint vectors in $Y \in \mathbb{R}^{n \times l}$, devices to
compute $A\ast X$, $B\ast X$ and $T\ast X$.\\
\> \textup{1.}\>Allocate memory  
$W,P,Q,AX,AW,AP,BX,BW,BP \in\mathbb{R}^{n \times m},
BY \in  \mathbb{R}^{n \times l}$. \\
\> \textup{2.} \> Apply the constraints to $X$: \\ 
\>\> $BY=B\ast Y$; $X=X-Y\ast \left( Y^T\ast BY\right)^{-1}\ast ((BY)^T \ast X)$. \\
\> \textup{3.} \>  $B$-orthonormalize $X$: $BX=B\ast X; R=\mbox{chol}(X^T\ast BX); 
X=X\ast R^{-1};$\\
\>\> $BX=BX\ast R^{-1}$; $AX=A\ast X$. 
(Note: ``$\mbox{chol}$'' is the Cholesky decomposition). \\
\> \textup{4.} \> Compute the initial Ritz vectors: solve the eigenproblem\\
\>\> $(X^T\ast AX)\ast TMP = TMP\ast \Lambda;$ \\
\>\> and compute $X=X\ast TMP; AX=AX\ast TMP$; $BX=BX\ast TMP$. \\ 
\> \textup{5.}  \> Define the index set $J$ of active iterates to be $\{1,\ldots ,m\}.$\\
\> \textup{6.}  \> {\bf for} $k=0,\ldots,MaxIterations$:\\
\> \textup{7.}\>\>Compute the residuals: $W_J=AX_J-BX_J\ast \Lambda_J$.\\
\> \textup{8.}\>\>Exclude from the index set $J$ the indices that correspond to residual \\
\>\>\> vectors for which the norm has become smaller than the tolerance.\\
\>\>\> If $J$ then becomes empty, exit loop. \\
\> \textup{9.}\>\>Apply the preconditioner $T$ to the residuals: $W_J=T\ast W_J$.\\
\> \textup{10}\>\>Apply the constraints to the preconditioned residuals $W_J$: \\
\>\>\> $W_J=W_J-Y\ast\left( Y^T\ast BY\right)^{-1}\ast ( (BY)^T \ast W_J )$.\\
\> \textup{11.}\>\>Compute $BW_J$ and $B$-orthonormalize $W_J$: $BW_J=B\ast W_J$;\\
\>\>\> $R=\mbox{chol}(W_J^T\ast BW_J)$; $W_J=W_J\ast R^{-1}$; $BW_J=BW_J\ast R^{-1}$. \\
\> \textup{12.}\>\>Compute $AW_J$: $AW_J=A\ast W_J$.\\
\> \textup{13.}\>\> {\bf if} $k>0$\\
\> \textup{14.}\>\>\>$B$-orthonormalize $P_J$: $R=\mbox{chol}(P_J^T\ast BP_J); P_J=P_J\ast R^{-1}$;\\
\> \textup{15.}\>\>\>Update $AP_J=AP_J\ast R^{-1}$; $BP_J=BP_J\ast R^{-1}$.\\
\> \textup{16.}\>\> {\bf end if}\\
\> \textup{   }\>\> {\bf Perform the Rayleigh Ritz Procedure:}\\
\> \textup{   }\>\>\> {\bf Compute symmetric Gram matrices:}\\   
\> \textup{17.}\>\>\> {\bf if} $k>0$\\
\> \textup{18.}\>\>\>\>$gramA=\left[\begin{array}{ccc}
\Lambda & X^T\ast AW_J & X^T\ast AP_J\\
\cdot & W_J^T\ast AW_J & W_J^T\ast AP_J\\
\cdot & \cdot & P_J^T\ast AP_J\\ \end{array}\right]$.\\ \\
\> \textup{19.}\>\>\>\>$gramB=\left[\begin{array}{ccc}
I & X^T\ast BW_J & X^T\ast BP_J\\
\cdot & I & W_J^T\ast BP_J\\
\cdot & \cdot & I\\
\end{array}\right]$.\\
\> \textup{20.}\>\>\> {\bf else}\\
\> \textup{21.}\>\>\>\>$gramA=\left[\begin{array}{cc}
\Lambda & X^T\ast AW_J\\
\cdot & W_J^T\ast AW_J\\
\end{array}\right]$.\\ \\
\> \textup{22.}\>\>\>\>$gramB=\left[\begin{array}{cc}
I & X^T\ast BW_J\\
\cdot & I\\ \end{array}\right]$.\\
\> \textup{23.}\>\>\> {\bf end if}\\
\> \textup{24.}\>\>\> {\bf Solve the generalized eigenvalue problem:}\\ 
\>\>\>\>$gramA \ast C = gramB\ast C \ast \Lambda$,
where the first $m$ eigenvalues in\\ 
\>\>\>\>increasing order are in the diagonal matrix $\Lambda$ and 
the \\
\>\>\>\>$gramB$-orthonormalized eigenvectors are the columns of $C$.\\
\> \textup{   }\>\>\> {\bf Compute Ritz vectors:}\\
\> \textup{25.}\>\>\> {\bf if} $k>0$\\
\> \textup{26.}\>\>\>\>Partition
$C=\left[\begin{array}{c}C_X\\C_W\\C_P\\\end{array}\right]$
according to the number of columns in\\
\>\>\>\>\>$X,~W_J$, and $P_J$, respectively.\\
\> \textup{27.}\>\>\>\>Compute $P=W_J\ast C_W+P_J\ast C_P$;\\
\>\>\>\>\>$AP=AW_J\ast C_W+AP_J\ast C_P$; $BP=BW_J\ast C_W+BP_J\ast C_P$.\\
\> \textup{28.}\>\>\>\>$X=X\ast C_X+P;AX=AX\ast C_X+AP; BX=BX\ast C_X+BP$.\\
\> \textup{29.}\>\>\> {\bf else}\\
\> \textup{30.}\>\>\>\>Partition
$C=\left[\begin{array}{c}C_X\\C_W\\\end{array}\right]$
according to the number of columns in\\
\>\>\>\>\>$X$ and $W_J$ respectively. \\
\> \textup{31.}\>\>\>\> $P=W_J\ast C_W; AP=AW_J\ast C_W; BP=BW_J\ast C_W$.\\
\> \textup{32.}\>\>\>\> $X=X\ast C_X+P;AX=AX\ast C_X+AP;BX=BX\ast C_X+BP$.\\
\> \textup{33.}\>\>\> {\bf end if}\\
\> \textup{37.} \> {\bf end for}\\
{\underline {\bf Output:}} The eigenvectors $X$ and the eigenvalues $\Lambda$.\\
\entab}

For description of all LOBPCG-BLOPEX variables see Appendix \ref{sec:de_va}.

In the complex case, the transpose needs to be replaced the adjoint. 
Only double-precision real arithmetic is supported in the current revision 1.0 of BLOPEX. 

The algorithm is \emph{matrix-free} in the sense that
the matrices $A$ and $B$ are not needed in the algorithm and not stored in the code, but 
rather are accessed only through matrix-vector product functions,  
provided by the user. Matrix-free codes are needed in applications 
where the matrices $A$ and $B$ are never explicitly formed, 
e.g.,\ in finite element method software packages for partial differential equations. 

The algorithm uses only one block application 
of the preconditioner $T$ (step 9)
and one block matrix-vector product $Bx$ (step 11) and $Ax$ (step 12), per iteration.  
This is achieved by storing and manipulating $9m$ vectors
($6m$ vectors if $B=I$).  
If there is not enough memory for $9m$ vectors, or 
if $m$ is so large that linear algebra operations with 
$9m$ vectors are more costly than multiplication by 
$A$ and $B$, e.g.,\ if $A$ and $B$ are very sparse,
the algorithm can be modified,  
so that only $3m$ vectors are stored, but then additional 
block matrix-vector products $Ax$ and $Bx$
are required per iteration. Such a modification is not yet 
currently available in BLOPEX. 

The issue of a good choice of the basis 
to use in the Rayleigh-Ritz method becomes even more delicate 
with a larger number $m$ of vectors in the block. 
Ill-conditioned Gram matrices in the Rayleigh-Ritz procedure  
may result in algorithm failure, inaccurate computations of the 
Ritz pairs, or spurious Ritz pairs. 
As it can be seen in the LOBPCG algorithm in steps 11-22 
we explicitly compute $X$, $P$, and $W$,   
and form the basis of the Rayleigh-Ritz procedure 
in the following way: $X$ is left untouched, 
since it has already $B$-orthonormal columns, made of the Ritz vectors, 
while the vectors inside both $W$ and $P$ are 
$B$-orthonormalized in steps 11 and 14, correspondingly, 
so we put the identities on the block diagonal of the matrix $gramB$    
(step 19/22). 

For the $B$-orthonormalization of $W$ and $P$ 
in steps 11 and 14, 
the cheapest orthonormalization technique, 
using the Cholesky decomposition 
of the Gram matrix, has been chosen. 
This technique is known to
be of a poor quality in the presence of the round-off errors.
Nevertheless, in our experience, 
the Cholesky-based $B$-orthonormalization of $W$ and $P$ reduces the risk of the 
Rayleigh-Ritz procedure failure while being considerably cheaper 
than the stable $B$-orthonormalization 
of all $3m$ vectors in the basis of the trial subspace 
described, e.g.,\ in \cite{hl06}. 

Let us note that matrices $P$ and $W$ converge to zero, 
even worse, different columns of $P$ and $W$ converge to zero
with different speeds, so matrices plugged in the 
Cholesky decomposition are extremely badly scaled. 
It is crucial for the stability that the code of the 
Cholesky decomposition used here is numerically 
invariant with respect to matrix scaling, otherwise, 
the explicit normalization of columns of $P$ and $W$ 
is necessary prior to the call of the Cholesky decomposition. 
Such a subtle property as scaling invariance  
that we rely upon here may apparently be affected 
by aggressive optimization flags during the compilation 
of the Cholesky decomposition code. 

We have also tested (in MATLAB) a version where the block 
diagonal of the matrix $gramB$ is explicitly computed instead of just 
setting the blocks to be the identities. Such an approach is similar to 
a known idea of repeated orthonormalization and results in a better 
attainable accuracy. However, in our tests we have found  
the attainable accuracy of our Algorithm \ref{slobpcg2} to be 
already adequate, so this is not implemented in our BLOPEX LOBPCG. 

Numerical software development  
is often an act of a balanced compromise between performance, reliability, and accuracy.
Performance has been our main priority in BLOPEX release 1.0, 
preserving reasonable reliability for common situations, and achieving the accuracy 
typical for large-scale applications in engineering. 
Design of other efficient and reliable implementations 
of the LOBPCG algorithm for BLOPEX is underway. 
\subsection{LOBPCG deflation by hard and soft locking} 
\label{slobpcg4}
When several eigenvectors are computed simultaneously, it is often 
the case that some eigenvectors converge faster than others. 
To avoid the unnecessary computational work, 
one ``locks'' the eigenvectors 
that have already converged within a required tolerance 
while continuing to iterate the others.
Locking approximately computed eigenvectors 
is not well studied theoretically.
In this subsection, we introduce hard and soft locking,  
and explain how locking is used in the LOBPCG algorithm, following 
the presentation of \citet{k_CM04}. 
The idea of soft locking is very natural and may have been known to some practitioners, 
developing codes for symmetric eigenvalue problems, even prior to \cite{k_CM04}, 
where the term ``soft locking'' has apparently been first introduced, but we are not aware 
of any earlier references.  

A commonly used locking technique,
known as the deflation by restriction, 
is to keep
the locked eigenvectors unchanged
and those
that are still iterated
orthogonal to the locked ones. 
This technique is usually called ``locking,'' but here it is 
referred to as ``hard locking'' to distinguish it from a new type of locking
described next.  

A different technique, called here ``soft locking'',  
can be used in
simultaneous iterations combined with the Rayleigh-Ritz procedure,
The idea of this technique is to remove the residuals of
the locked vectors from the computation,
but continue
using the vectors themselves in the Rayleigh-Ritz procedure,
which can change them.
The orthogonality of the soft locked vectors to iterative vectors 
follows from the well-known fact that the Ritz vectors 
corresponding to different Ritz values are orthogonal,
and those corresponding to multiple Ritz values
can be chosen to be orthogonal.  

Compared to the hard locking,
soft locking is computationally more expensive. 
The advantage of the soft locking is that 
it provides more accurate results. 
First and foremost, as the number of hard locked vectors increases,
the attainable accuracy of the iterated ones may decrease, 
which may prevent them from reaching the required tolerance.
With soft locking, the attainable accuracy of the iterated vectors
does not depend on the number and the accuracy of soft locked ones.
Second, if the Rayleigh quotient of $k$-th locked vector is less than
the $(k+1)$-th exact eigenvalue, then
the accuracy of the first $k$ locked vectors will increase
in the course of the iterations, owing to the participation 
in the Rayleigh-Ritz procedure, even though their role in
the iterative loop is now reduced to this procedure only.

Below we describe the hard and soft locking 
as used in the BLOPEX implementation of LOBPCG
and present preliminary  
numerical results demonstrating the effectiveness of the soft locking 
compared to the traditional hard locking. 

Let us consider the single-vector LOBPCG method with no locking
as described by (\ref{old}) 
$
x^{(i+1)} = w^{(i)} + \tau^{(i)} x^{(i)} + \gamma^{(i)} x^{(i-1)}, \,
w^{(i)} =  T (A x^{(i)} - \lambda^{(i)} B x^{(i)})
$
that we repeat here for the reader's convenience. 
Let $P$ be a projector on the complement to the span of 
already computed eigenvectors. 
In LOBPCG with the hard locking, we redefine
$w^{(i)} =  {P} T (A x^{(i)} - \lambda^{(i)} B x^{(i)}),$
so that all iterates are now within the range of $P$ 
if the initial guess in chosen within the range. 
Traditionally, $P$ is orthogonal in the $B$-based scalar 
product and the complement is orthogonal also in the 
$B$-based scalar product, since this allows the easiest 
and most inexpensive implementation. 
This is what we use in step 10 of the 
LOBPCG algorithm in Subsection \ref{slobpcg2}.
The constraints, given by the columns of the matrix $Y$,  
are in this case the already approximately computed 
eigenvectors that we want to deflate. 

If the locked eigenvectors are computed exactly 
then it is easy to analyze the influence of classical 
hard locking and to show that it is equivalent to 
simply removing the locked eigenpairs from the consideration. 
If the locked  eigenvectors are only approximate, 
the range of P is not exactly an invariant subspace
of the matrix pencil  $A-\lambda B$, and   
the analysis of the influence of the approximation quality 
on locking becomes much more complicated; 
see e.g. \citet{dk92}. 

As the number of locked vectors increases, their inaccuracy  
may affect the attainable accuracy of
iterated vectors. 
To avoid this,
in the LOBPCG algorithm in Subsection \ref{slobpcg2}
we use soft locking.
We denote by $X_J$ the matrix that contains 
only the active (not soft locked) 
iterative eigenvector approximations as columns, while all 
$m$ iterative vectors $X$ are included in the Rayleigh-Ritz 
procedure. 
Only $X_J$ participates in iterations, i.e., having  
$\Lambda_J^{(i)}$ and $X_J^{(i)}$ we compute only
$W_J^{(i)}=T(A X_J^{(i)} - B X_J^{(i)} \Lambda_J^{(i)})$,   
where  $\Lambda_J^{(i)}$ is a diagonal matrix of approximate 
eigenvalues corresponding to  $X_J^{(i)}$. 
We include the complete $X$, but 
only $W_J$ and $P_J$, in the basis of the  
the Rayleigh-Ritz procedure 
in steps 18-22 and 26-32 
of the LOBPCG algorithm in Subsection \ref{slobpcg2}. 

\begin{figure}[!htb]
\begin{center}
\includegraphics[width=2.5in]{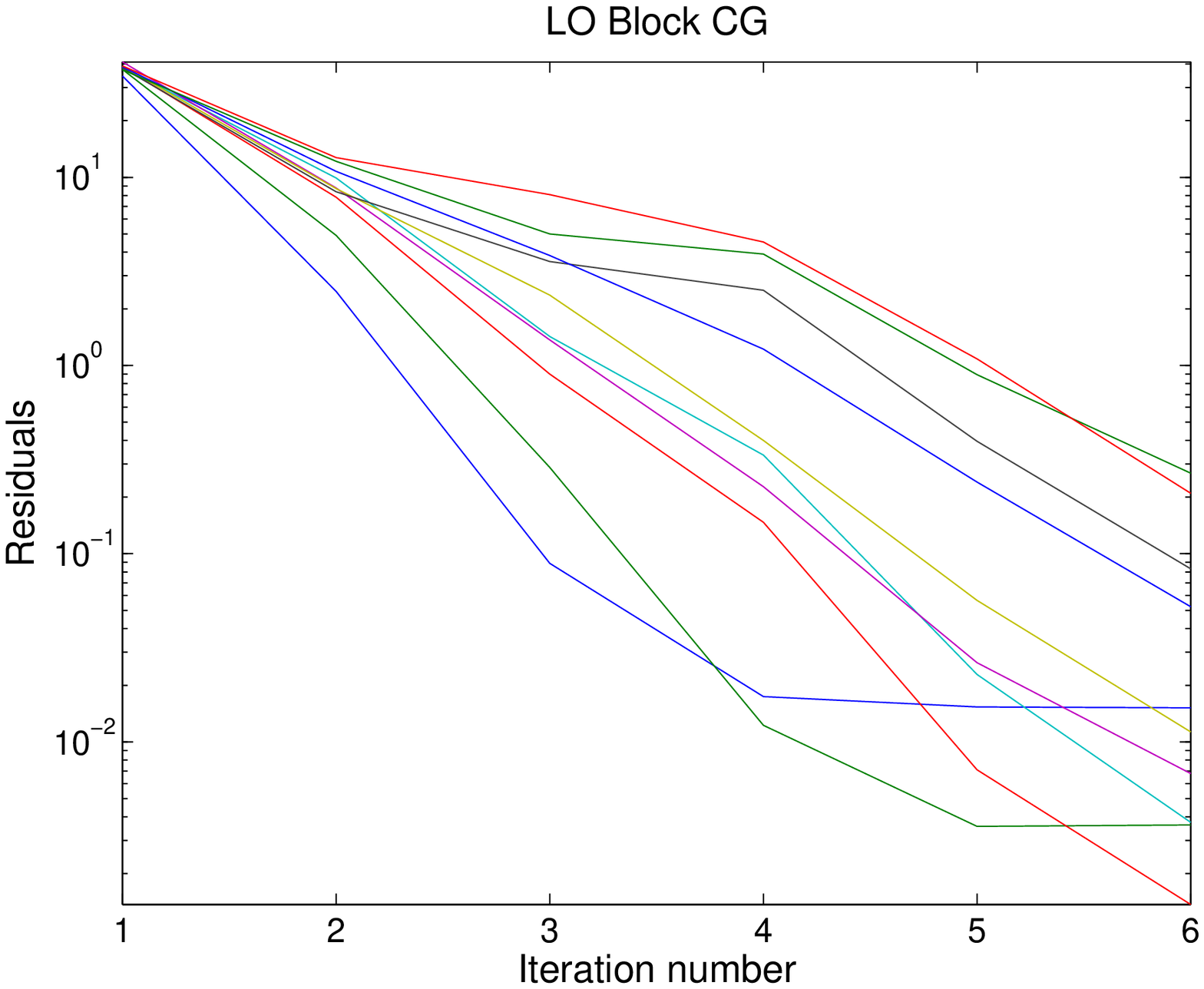}
\includegraphics[width=2.5in]{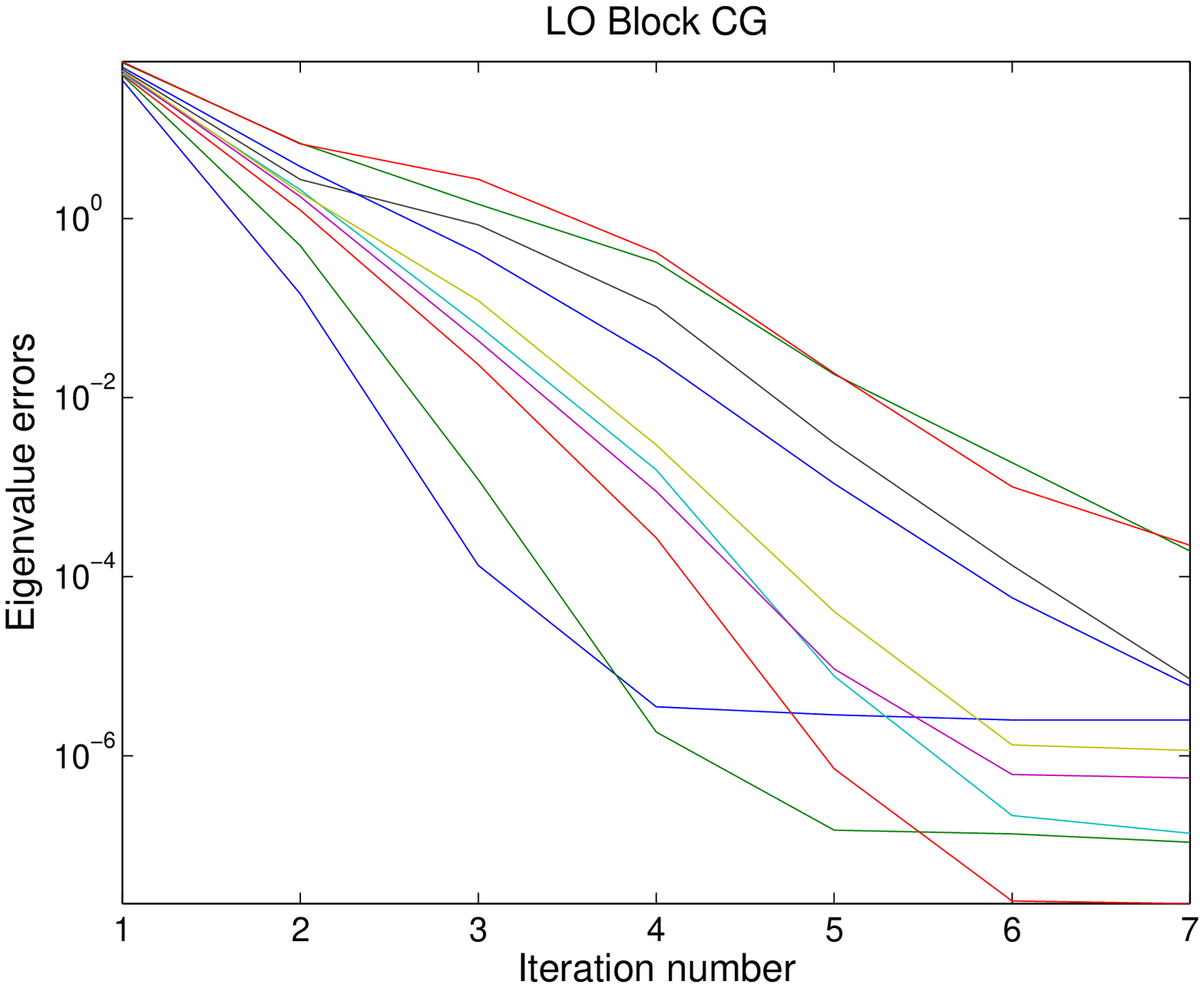}
\end{center}
\caption{Residual norms (left) and eigenvalue errors (right) 
for $10$ eigenpairs in LOBPCG with soft locking. Soft locked (the bottom 6) 
eigenpairs continue to improve.}
\label{f_sl}
\end{figure}

The positive influence of the 
soft locking is especially noticeable 
when the stopping tolerance is not so small. 
Figure \ref{f_sl} presents the results of the 
soft locking when the residual norm reaches $10^{-1}$
in LOBPCG with $m=10$ for a model problem. 
We observe that the active eigenpairs continue 
to converge and that the eigenpairs, which are already soft locked, still 
improve their accuracy. 

To summarize, we use the classical 
``hard'' locking in LOBPCG through the constraints, and 
soft locking  inside the LOBPCG code. If the user wants 
to compute so many eigenpairs that 
it is not reasonable to include all of them at once 
into the block, the user should determine 
an appropriate block size $m$ for the given hardware
and call LOBPCG with the block size $m$
several times, putting the previously computed eigenvectors into the 
constraints for the next LOBPCG call. The LOBPCG MATLAB code 
includes an example of such a call in the help information.  
\section{Abstract BLOPEX implementation for PETSc and {\it hypre}}
\label{sabs}
PETSC and {\it hypre} are 
software libraries aimed at solving large systems
on massively parallel computers. 
Our native PETSc BLOPEX version gives the PETSc user community easy access to 
the customizable code of a modern preconditioned eigensolver
and an opportunity to easily call {\it hypre} preconditioners from PETSc.  
The BLOPEX built-in {\it hypre} version efficiently takes direct advantage of powerful {\it hypre} 
AMG preconditioner BoomerAMG and its geometric multigrid preconditioners PFMG and SMG. 

\begin{figure}[ht]
\begin{center}\setlength{\unitlength}{1in}
\begin{picture}(5,1.8)
\put(.3,1.5){\framebox(1.8,.3){PETSc driver for LOBPCG}}
\put(2.5,1.5){\framebox(1.8,.3){{\it hypre} driver for LOBPCG}}
\put(.3,1){\framebox(1.8,.3) {Interface PETSc-BLOPEX}}
\put(2.5,1){\framebox(1.8,.3){Interface {\it hypre}-BLOPEX}}
\put(.1,.5){\framebox(1.3,.3){PETSc libraries}}
\put(1.5,.5){\framebox(1.6,.3){Abstract BLOPEX}}
\put(3.2,.5){\framebox(1.3,.3){{\it hypre} libraries}}
\put(1.2,1.5){\vector(0,-1){.2}}
\put(1.2,1.3){\vector(0,1){.2}}
\put(3.4,1.5){\vector(0,-1){.2}}
\put(3.4,1.3){\vector(0,1){.2}}
\put(0.7,1){\vector(0,-1){.2}}
\put(0.7,0.8){\vector(0,1){.2}}
\put(1.8,1){\vector(0,-1){.2}}
\put(1.8,0.8){\vector(0,1){.2}}
\put(2.9,1){\vector(0,-1){.2}}
\put(2.9,0.8){\vector(0,1){.2}}
\put(3.9,1){\vector(0,-1){.2}}
\put(3.9,0.8){\vector(0,1){.2}}
\end{picture}
\vspace{-1cm}
\end{center}
\caption{BLOPEX {\it hypre} and PETSc software modules}
\label{lobpcg_c_modules}
\end{figure}
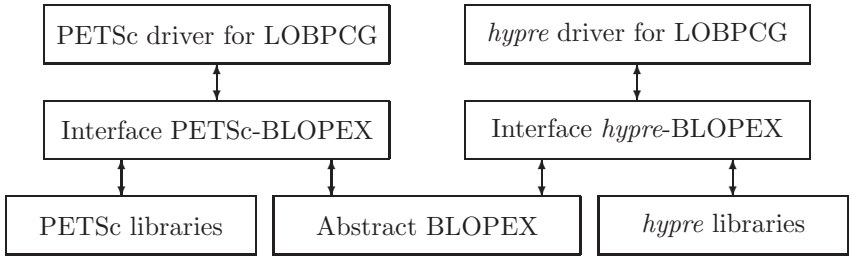

The BLOPEX library at present includes only the 
LOBPCG eigensolver. The BLOPEX code is written in C-language 
and calls a few LAPACK subroutines for dense matrix operations. 
The matrix-vector multiply and the preconditioner call are done
through user supplied functions. The main LOBPCG code is abstract in the sense 
that it works only through an interface that determines the particular 
software environment: PETSc or {\it hypre} or user defined, in order to call 
parallel (multi)vector manipulation routines. 
A block diagram of the main modules is given
in Figure \ref{lobpcg_c_modules}.

In the abstract part of BLOPEX, the blocks of vectors ($X$, $P$ etc) are represented by
an abstract data object we call ``multivector.'' 
A particular implementation of the multivector 
is outside of our code, so we only provide an interface to it. 
Ideally, all operations involving 
multivectors should be implemented in a block 
fashion, to minimize the number of exchanges between the 
processors and to take advantage of 
a highly optimized matrix-matrix multiplication
routine {\tt dgemm} in BLAS
when computing Gram matrices and updating multivectors. 
However, in PETSc and {\it hypre}
such an object is currently not available to users to our knowledge, so we 
had to temporarily use the multivector interfaces as references to individual vectors. 
When properly implemented multivectors become available in PETSc and {\it hypre}, 
we plan to change our interface codes to use the true multivectors, 
which is expected to lead to a great speed-up 
due to a smaller number of communications and efficient BLAS use.  
\section
{BLOPEX LOBPCG Numerical Results in PETSc and {\it hypre}}\label{snum}
Here we presents results accumulated over considerable length of time. 
Despite our efforts, the reader may still notice some inconsistency, e.g.,\
some tests may use different tolerances. All tests are performed in 
double precision real arithmetic. 
\subsection{The Basic Accuracy of Algorithm}\label{ss1-1}
In these tests BLOPEX LOBPCG computes the smallest $50$ eigenvalues of 
3D 7-point $200 \times 200 \times 200$ and $200 \times 201 \times 202$ 
Laplacians using a direct application of the {\it hypre} BoomerAMG preconditioning. 
In the first case we have eigenvalues with
multiplicity and in the second case the eigenvalues are distinct, but clustered.
The initial approximations are chosen randomly.
We set the stopping tolerance (the norm of the maximum residual)
equal to $10^{-6}$. 
The numerical output (for the complete data, see \cite{ka05tr}) 
is compared to the exact eigenvalues 
$\lambda_{i,j,k}=4\left[\sin\left(\frac{i\pi}{2(n_x+1)}\right)^2+
\sin\left(\frac{j\pi}{2(n_y+1)}\right)^2 + \sin\left(\frac{k\pi}{2(n_z+1)}\right)^2\right],$
of the 3D 7-point Laplacians on the 
$n_x \times n_y \times n_z$ cube
with $1 \le i \le n_x$, $1 \le j \le n_y$ and $1 \le k \le n_z$ 
In both tests for all eigenvalues the maximum
relative error is less than $10^{-8}$, so LOBPCG is cluster robust, i.e. 
it does not miss (nearly) multiple eigenvalues, which supports the 
conclusion of \citet{MR2199542}.  
\subsection{Performance Versus the Number of Inner Iterations}\label{ss1}
We can execute a preconditioner
$x=Tb$ directly or by calling PCG to solve a linear system, 
in our tests, $Ax=b$, so $T$ approximates $A^{-1}$.
Therefore, we can expect that increasing the number of ``inner''
iterations of the PCG might accelerate the overall  convergence, 
but only if we do not make too many iterations, since even if 
$T = A^{-1}$, the convergence of the LOBPCG method is still linear. 
In other words, for a given matrix $A$ and 
a particular choice of a preconditioner $T$,
there should be an optimal finite (may be zero, corresponding to the 
direct application of $T$) number of inner iterations.

We try to find this optimal number for the Schwarz-PCG
and BoomerAMG-PCG preconditioners in {\it hypre} and PETSc for 
the $7$-point $3$-D  $100\times100\times 100$ Laplacian
with  the block-size $m=1$. 
We measure the execution time as we
vary the quality of the preconditioner by changing the maximum number of
inner iterations in the corresponding PCG solver. 
PETSc and {\it hypre} built-in Schwarz algorithms 
are different, but demonstrate a similar behavior.  
The {\it hypre} multigrid BoomerAMG and PFMG preconditioners 
are called from {\it hypre} test 
drivers. The {\it hypre} BoomerAMG preconditioner in addition is called through PETSc. 
The results for these tests involving different multigrid preconditioners are also similar.  

We find that for this problem the optimal number of inner iterations is
approximately $10$--$15$ for Schwarz-PCG, but BoomerAMG-PCG works best if BoomerAMG 
is applied directly as a preconditioner, without even initializing the 
PCG solve function. 
Our explanation of this behavior is based on two 
facts. First, the Schwarz method with the default parameters used here 
is somewhat cheaper, but not of such a good 
quality, compared to BoomerAMG in these tests. Moreover, 
the costs for matrix-vector and multi-vector linear algebra in LOBPCG are  
relatively small compared to the costs of the BoomerAMG application, but is  
comparable to the costs of application of the Schwarz preconditioner here.
Second, one PCG iteration 
is less computationally expensive compared to one LOBPCG iteration 
because of larger number of linear algebra operations with vectors 
in the latter. A single direct application of BoomerAMG as the preconditioner 
in LOBPCG gives enough improvement in convergence to make it the best 
choice, while Schwarz requires more iterations that are less time consuming 
if performed using PCG, rather than by direct application in LOBPCG. 

In the next set of numerical tests for the same eigenproblem 
we use the block-size $m=10$, and we vary the
type of PCG preconditioner available in the  {\it hypre} IJ interface 
together with the number of the 
inner PCG iterations. For the complete test data, see \citet{ka05tr}; here we 
provide only the summary. 
We observe that both the number of outer iterations
and the execution time can be decreased significantly by choosing the maximum
number of inner iterations optimally. 
For this problem, the BoomerAMG preconditioner applied directly 
has the fastest run time and converges in the smallest number of iterations, $31$. 
For comparison, if we use no preconditioning at all then 
in $500$ iterations that take ten times longer 
we still do not reach the required tolerance $10^{-10},$ which 
illustrates the importance of preconditioning. 
\subsection{LOBPCG Performance vs. Block Size}
\begin{figure}[htb!]
\begin{center}
\includegraphics[width=3in]{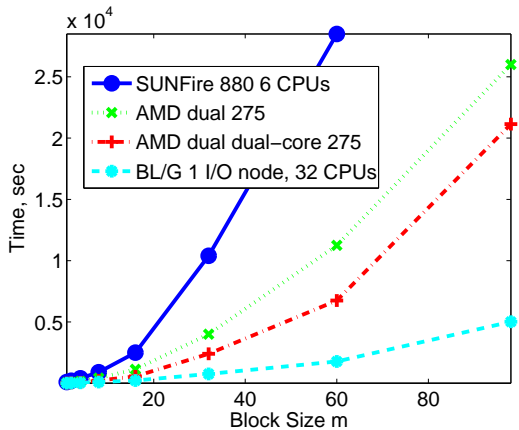}
\end{center}
\caption{LOBPCG Performance vs. Block Size. Preconditioner:  {\it hypre} BoomerAMG.} \label{fig_pb}
\end{figure}

We test {\it hypre} and PETSc LOBPCG on a 
$7$-point $3$D Laplacian with $128\times128\times128\approx 2M$ unknowns 
using the {\it hypre} BoomerAMG preconditioner 
by increasing the block size $m$ 
from $1$ to $98.$ 
We run the tests on shared memory SUN Fire880 system,  
Dual and Single Core AMD Opterons Dual 275 CPUs servers, and 
a single I/O node of IBM BG/L with $32$ CPUs. 
We find that the  {\it hypre} and PETSc LOBPCG run times are very close, 
using the same tolerance  $10^{-8}$, 
with PETSc extra overhead less than a few percents of the total times 
compared to a direct use of {\it hypre}, as expected.  
The growth of the CPU time 
with the increase of the block size $m$ is approximately linear for this 
problem up to $m=16$, then the complexity term $m^2 n$ becomes visible 
as well as the slower convergence rate for larger eigenvalues 
starts noticeably affecting the overall timing on Figure \ref{fig_pb}. 
\subsection{Scalability}\label{ss3}
For the $7$-point $3$D Laplacian we vary the problem size
proportional to the number of processors. 
We directly apply {\it hypre} BoomerAMG or PFMG multigrid preconditioners. 
First, on the LLNL MCR cluster 
the eightfold increase of the number of dual-CPU nodes does not noticeably increase the 
LOBPCG CPU time, which is approximately 50 sec for BoomerAMG and 180 sec for PFMG 
with $m=1$ and  tolerance  $10^{-8}.$
\begin{figure}[htb]
\begin{center}
\includegraphics[width=2.5in]{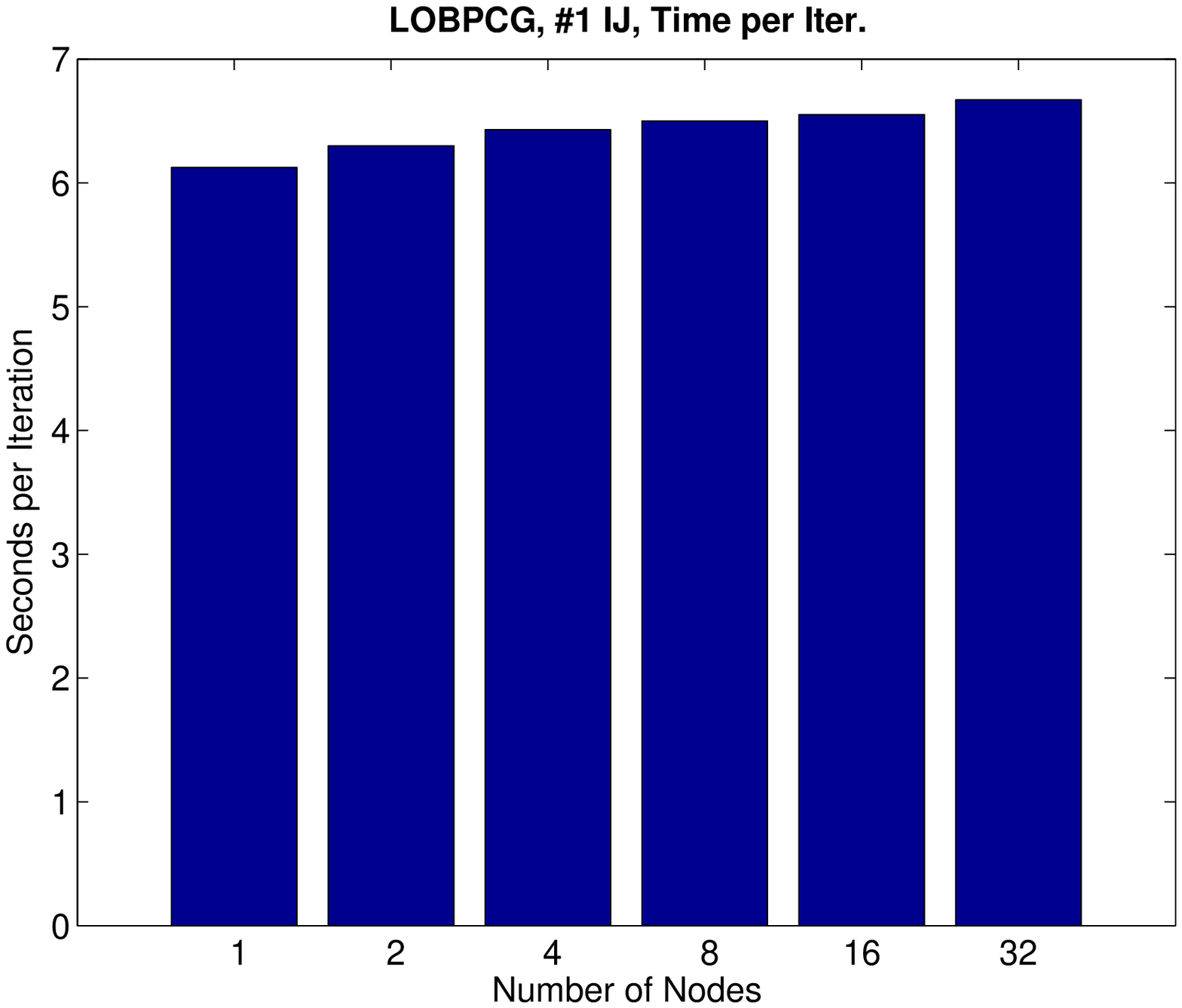}
\includegraphics[width=2.5in]{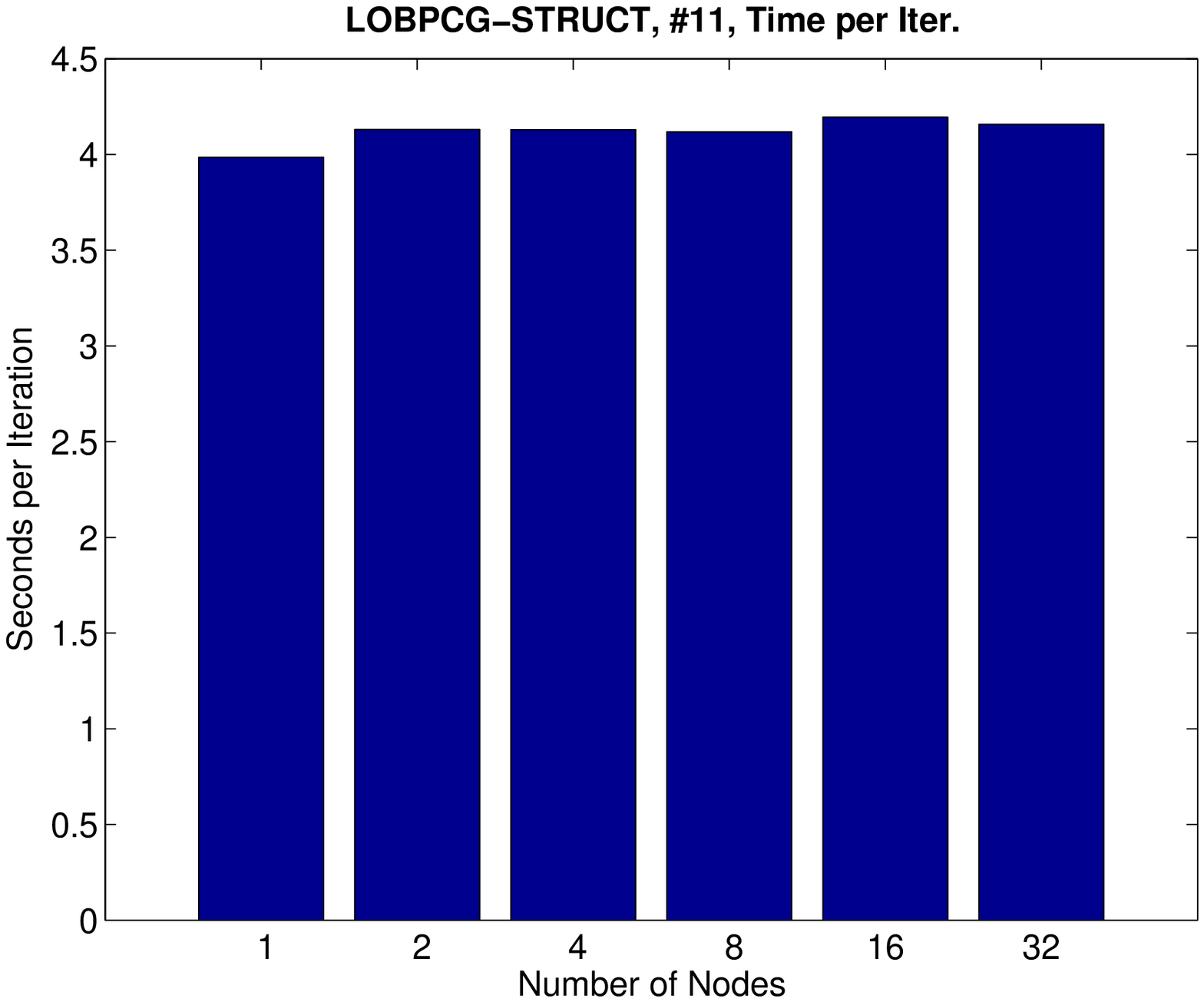}
\end{center}
\caption{LOBPCG scalability, 1M unknowns per node on Beowulf.}\label{f:mg2}
\end{figure} 
The PFMG takes more time overall compared to BoomerAMG because of the larger convergence factor, 
but each iteration of the former is 50\% faster than that of the latter, 
as we observe on Figure \ref{f:mg2}, which displays the CPU time per iteration 
for LOBPCG {\it hypre} code with {\it hypre} BoomerAMG and PFMG preconditioners 
in our second test, on the Beowulf cluster. 
We see that the 32-fold increase in the number of dual-CPU nodes just slightly increases 
the LOBPCG CPU time.

\begin{table}
\begin{center}
\begin{tabular}{|r|r|c|c|c|}\hline
\# CPUs&Matrix Size&Setup (sec) &\# Iter. &Solve time (sec.)\\\hline
8&   4.096   M&         7&    9 / 19 &     62 / 580 \\ \hline
64&  32.768  M&         10&   8 / 17 &     62 / 570 \\ \hline
512& 0.262144 B&        15&   7 / 13 &     56 / 500 \\ \hline
\end{tabular}
\end{center}
\caption{Scalability for $80\times80\times80=512,000$ mesh per CPU, 
$m=$ 1 / 6.} \label{g6_table}
\end{table}
\begin{table}
\begin{center}
\begin{tabular}{|r|r|c|c|c|}\hline
\# CPUs&Matrix Size&Setup (sec)& \# Iter.&Solve time (thous. sec.)\\\hline
8  &   1 M&  1&   34&    3\\\hline
64 &   8 M&  5&   32&    3\\ \hline
512&  64 M& 10&   29&    3\\ \hline
\end{tabular}
\end{center}
\caption{Scalability for $50\times50\times50=125,000$ mesh per CPU, 
$m=50$.} \label{g7_table}
\end{table}
Finally, we solve the $7$-point $3$D
Laplacian eigenvalue problems for large matrix sizes and for 
a variety of block sizes $m$ on the BG/L system. We observe a $10$--$20$\% 
variability in the iteration numbers and thus in the wall clock, 
due to the randomness of the initial approximations, 
so we report here typical average results.  
Tables \ref{g6_table} and {\ref{g7_table} 
contain the information about LOBPCG runs using  
{\it hypre} SMG geometric multigrid and tolerance $10^{-6}$.  
\section*{Conclusions}
The BLOPEX software has been integrated into
the {\it hypre} library and is easily available in PETSc 
as an external package. 
BLOPEX is the only currently available software that 
solves eigenvalue problems using high quality existing  
{\it hypre} and PETSc preconditioners directly.
Our abstract C implementation of the LOBPCG in BLOPEX allows 
easy deployment with different software packages.  
Our interface routines are based on {\it hypre} and 
PETSc standard interfaces and give the user an opportunity 
to provide matrix-vector multiply and preconditioned solver functions.
The code scalability is tested for model problems of large sizes on 
parallel systems. 
\section*{Acknowledgments}
We are grateful to all members of the Scalable Algorithms
Group of the Center for Applied Scientific Computing,
Lawrence Livermore National Laboratory and, in particular, to Ulrike Yang, Rob Falgout,
Edmond Chow, Charles Tong, and Panayot Vassilevski, for their patient support
and help.
We thank Matthew Knepley and other members of the 
PETSc team for their help with incorporating our BLOPEX as 
an external package into the PETSc library. 
We thank Jose Roman, a member of SLEPc team, for writing the SLEPc interface to our  {\it hypre} LOBPCG.  


\def\cprime{$'$}

\appendix
\section{Description of all LOBPCG-BLOPEX variables}\label{sec:de_va}
\begin{description}
\item[$X$] (multivector). The iterative approximation to eigenvectors. 
\item[$A$] (function). Multiplies the function argument, which is a multivector, by the first 
matrix of the eigenproblem. 
\item[$B$] (function). Multiplies the function argument, which is a multivector, by the second 
matrix of the eigenproblem. 
\item[$T$] (function). Applies the preconditioner to the function argument, a multivector. 
\item[$Y$] (multivector). The given constraint that can be used for the hard locking. 
Iterations run in the $B$-orthogonal complement to the span of $Y$.
\item[$\Lambda$] (diagonal matrix, in the actual code given by a vector of the diagonal entries). 
The iterative approximation to eigenvalues.
\item[$W$] (multivector). Used to store the residuals (step 7) and 
the preconditioned  residuals (step 9) and the preconditioned constrained residuals (step 10 and later). 
\item[$J$] (index vector). Denotes the set of indexes of active iterates. 
The starting value for $J$ is $\{1,\ldots ,m\}$. 
If the norm of the residual $r_j$ for some $j \in J$ 
has become smaller than the tolerance, this index $j$ is  
excluded from the set $J$. By $X_J$ we denote 
a sub-matrix of $X$ with the columns, which indexes are present in 
the set $J$---these are the active iterates. 
The  indexes from $\{1,\ldots ,m\}$ that are not 
present in the set $J$ correspond to iterative 
approximations to eigenvectors, which have been ``soft locked.'' 
\item[$P$] (multivector). The LOBPCG ``directions.''
\item[$gramA$ and $gramB$] (matrices). The Gram matrices in the Rayleigh-Ritz procedure. 
\item[$C$]  (matrix). The matrix of the $gramB$-orthonormalized eigenvectors 
in the Rayleigh-Ritz procedure, used as the matrix of coefficients 
to compute the Ritz vectors, the new multivectors $X$, 
as linear combinations of the basis $X, W_J,$ and $P_J$.  
\item[$AX, BX, AP, BP$ etc.] (multivectors). Implicitly computed results of the 
application of the corresponding functions to the corresponding vectors. 
In the absence of the round-off errors would be identical to their targets,
e.g.,\ $AX=A\star X$. 
\item[$R$]  (matrix). The upper triangular matrix of the Cholesky factorization, 
used for the $B$-orthogonalization via Cholesky. 
\end{description}
\section{Installing and testing BLOPEX with PETSc and {\it hypre}}
\label{sec:in_te}
Here we give brief specific information 
for the reader who wants to install and test our BLOPEX software. 
This information concerns the current BLOPEX-1.0,  
PETSc-2.3.2, and {\it hypre}-2.0.0 releases. We refer to Appendix \ref{sec:acro}
for the acronyms and other names not explained here and suggest reading the 
software manuals for more details. 
 
To install and test BLOPEX and {\it hypre} in PETSc, download the PETSc source  
and follow the standard installation procedure with the external packages option. 
The present procedure is to add 
``-{}-download-blopex=1 -{}-download-hypre=1'' options to the PETSc 
configuration. BLOPEX and {\it hypre} sources are then downloaded 
and compiled automatically. The BLOPEX test driver is already 
included in PETSc at the src/contrib/blopex/driver directory. 
In the following example:
\\
driver  -da\_grid\_x 128  -da\_grid\_y 128  -da\_grid\_z 128 -ksp\_type preonly -pc\_type hypre -pc\_hypre\_type boomeramg -n\_eigs 10 -seed 2 -tol 1e-6 -itr 5 -full\_out 1 
\\
the eigenproblem for the $7$-point $3$D Laplacian is solved using  
the {\it hypre} BoomerAMG preconditioner directly. 
The last five parameters for BLOPEX are self-explanatory. 

BLOPEX LOBPCG is built into {\it hypre} and requires no special 
configuration options. 
{\it hypre} supports four conceptual interfaces: STRUCT, SSTRUCT, FEM and IJ.
LOBPCG  has been tested with all but the FEM interface.
{\it hypre}-2.0.0 test drivers for LOBPCG are in the src/test directory. 
Examples with the same options as in the PETSc call above using IJ and STRUCT test drivers: 
\\
ij -lobpcg  -n 128 128 128 -pcgitr 0 -vrand 10 -seed 2 -tol 1e-6 -itr 5 -verb 1 
\\
struct -lobpcg  -n 128 128 128 -pcgitr 0 -vrand 10 -seed 2 -tol 1e-6 -itr 5 -verb 1

Our code can, but is not really intended to, be used with 
the popular shift-and-invert strategy, since its main advantage 
compared to traditional eigensolvers is that it uses preconditioning. 
Our test drivers can call preconditioning directly
(``-ksp\_type preonly'' PETSc option and ``-pcgitr 0'' {\it hypre} option) 
as well as through calls
to the {\it hypre}/PETSc PCG method. In the latter case 
the action $x = T b$ of the preconditioner
$T$ on a given vector $b$ is performed by 
calling a few steps of PCG to solve $A x = b$.
The BLOPEX LOBPCG code has been tested with all available {\it hypre} PCG-capable preconditioners 
in STRUCT, SSTRUCT and IJ interfaces, with the PETSc
native Additive Schwarz, and with the PETSc linked BoomerAMG from {\it hypre}. 
\section{Computers used for testing}
\label{sec:comp}
\begin{description}
\item[AMD workstations]
Several single and dual core AMD two 64 bit 
275 CPUs with 16GB shared memory, running Linux, at CCM CU-Denver. 
\item[SUN Fire 880] SUN Solaris 9, 6 UltraSPARC III 64 bit CPUs, 24GB shared memory, at CCM CU-Denver. 
\item[Beowulf cluster] Linux, 
36 nodes, dual PIII 933MHz processors and 2GB memory per node with 7.2SCI Dolphin interconnect, at CCM CU-Denver. 
\item[MCR cluster] Linux, dual Xeon 2.4GHz 4 GB nodes with Quadrics, at LLNL. 
\item[Blue Gene/L  (BG/L)] IBM single-rack with 1024 compute nodes, organized in 32 I/O nodes with 32 compute nodes each. Every compute node is a dual-core chip, containing two 700MHz PowerPC-440 CPUs and 512MB of memory. 
We run using the default 1 CPU for computing and 1 CPU for communication 
on every compute node. 
Location: NCAR.
\end{description}
\section{Acronyms and other names}\label{sec:acro}
\subsection{Organizations} 
\begin{description}
\item[CCM CU-Denver]
Center for Computational Mathematics, 
University of Colorado at Denver and Health Sciences Center, 
Downtown Denver Campus.
\item[LLNL] Lawrence Livermore National Laboratory.
\item[NCAR] National Center for Atmospheric Research. 
\end{description}
\subsection{Software packages} 
\begin{description}
\item[BLOPEX]
Block Locally Optimal Preconditioned Eigenvalue Xolvers, 
a library for large scale eigenvalue problems, developed 
by the authors of the paper. 
\item[PETSc]
``Portable, Extensible Toolkit for Scientific Computation,''
tools for the scalable solution of partial differential equations and related problems
developed by Argonne National Laboratory, see \citet{petsc-user-ref}.
\item[{\em hypre}]
``High Performance Preconditioners,''
scalable linear solvers package developed by LLNL, see \citet{Falgout:2005:PSH,Falgout:2006}.
\item[SLEPc]
``Scalable Library for Eigenvalue Problem Computations,''
a library for large eigenvalue problems, an extension of PETSc, 
see \citet{Hernandez:2005:SSF}. 
\end{description}
\subsection{{\em hypre} interfaces and preconditioners} 
\begin{description}
\item[IJ] 
A Linear-Algebraic Interface for applications with sparse matrices. 
\begin{description}
\item[BoomerAMG] a parallel implementation of an algebraic multigrid. 
\item[Schwarz]
Agglomeration-based domain decomposition preconditioner.  
\end{description}
\item[STRUCT] 
A structured grid interface called that aims applications with logically rectangular grids.
\begin{description}
\item[SMG]
A parallel semicoarsening multigrid solver for the linear systems arising from finite difference,
finite volume, or finite element discretizations of the diffusion equation 
on logically rectangular grids. The code solves both 2D and 3D problems with discretization stencils
of up to 9-point in 2D and up to 27-point in 3D. 
The algorithm semicoarsens in the z-direction and uses plane smoothing. 
\item[PFMG]
PFMG is a parallel semicoarsening multigrid solver similar to SMG. 
The main difference between the two methods is in the smoother: PFMG uses simple pointwise
smoothing. As a result, PFMG is not as robust as SMG, but is much more efficient per V-cycle.
\end{description}
\item[SSTRUCT] 
A semi-structured grid interface named that targets applications with grids that are mostly structured, but with some unstructured features. 
\item[FEI]
An unstructured grid interface designed for Finite Element applications. 
\end{description}
\subsection{Methods} 
\begin{description}
\item[PCG] Preconditioned Conjugate Gradient.
\item[LOBPCG] Locally Optimal Block Preconditioned Conjugate Gradient.
\item[AMG] Algebraic multigrid. 
\end{description}
\end{document}